\documentclass{article}

\usepackage{iclr2027_conference,times}  % official ICLR 2027 kit, UNMODIFIED
\iclrfinalcopy  % authors are printed in this build

\usepackage[utf8]{inputenc}
\usepackage[T1]{fontenc}
\usepackage{hyperref}
\usepackage{url}
\usepackage{booktabs}
\usepackage{amsfonts}
\usepackage{amsmath}
\usepackage{amssymb}
\usepackage{nicefrac}
\usepackage{microtype}
\usepackage{xcolor}
\usepackage{enumitem}
\usepackage{tikz}
\usetikzlibrary{arrows.meta, positioning}
\usepackage{pgfplots}
\pgfplotsset{compat=1.18}

\title{Guiding the Coarse Levels of Semantic IDs\\ Makes the Fine Levels Learnable}

\author{Bin Wang \& Zhengyu Zhang \\
Meta \\
\texttt{\{binwang88,zhengyuzhang\}@meta.com}
}

\begin{document}

\maketitle
% The style file sets \lhead inside \@maketitle, so the banner can only be cleared
% AFTER \maketitle. The paper is not published, so it must not claim to be.
\lhead{}  % or \lhead{Preprint.} for a running head
% The manuscript. Single copy, shared by both builds -- this is the file to edit.
% Runs from the abstract through the appendices; the drivers supply the document
% environment and the title around it.

\begin{abstract}
Generative retrieval represents each item by a short \emph{Semantic ID (SID)}---a sequence of
discrete codes from a residual-quantized autoencoder (RQ-VAE)---and casts recommendation as
autoregressive generation of that sequence. Because the tokenizer is trained \emph{independently} to
reconstruct an item embedding, its codes are aligned with neither the downstream LLM (they enter the
vocabulary as opaque tokens) nor the end task. Nearly every SID system therefore spends extra effort
to bridge this gap---alignment corpora, reasoning/RL, or per-token encoders to make codes
\emph{legible}, or learned tokenizer supervision to make them \emph{task-aware}---yet the recovered
meaning is content-derived and may not be the meaning the task needs. We introduce \textbf{Guided
SID}, which instead makes the levels that matter most meaningful \emph{by construction}: we
\emph{force} the coarse RQ-VAE levels to encode a predefined categorical attribute---chosen to be
text-grounded (hence legible to the LLM) \emph{and} task-relevant---by
deterministic supervised index assignment (overriding nearest-neighbor selection with the
attribute label) while keeping the codebooks learnable (they still receive reconstruction
gradients). A trie-merge construction maps any high-cardinality or set-valued attribute onto
the fixed code budget while keeping merged buckets semantically coherent. Guiding costs nothing
intrinsically: collision and reconstruction match or beat the vanilla baseline despite
pinning the coarse level. In a matched end-to-end A/B differing \emph{only} in the SID
encoding, the guided retriever improves recall@$k$ at every list length we measure
($1.36\times$ at $k{=}1$, $1.39\times$ at $k{=}10$) against the full production catalog of
654--741M identifiers, raises mean reciprocal rank from $0.0260$ to $0.0355$, and predicts the
pre-defined attribute $4.2\times$ more often. A third arm that instead
\emph{prepends} the attribute as an extra token, leaving the content codes untouched, recovers
almost none of that gain---which locates the effect in the \emph{restructured coarse code} rather
than in conditioning on the attribute. Guiding also lifts the \emph{free residual} codebooks it never supervises---$1.4$--$1.5\times$ the
baseline at every level---so a supervised coarse partition makes the finer, unsupervised levels
easier for an autoregressive model to generate, though by a far smaller margin than it improves the
coarse level itself.
\end{abstract}

\section{Introduction}

A now-standard paradigm for large-scale recommendation replaces an item's atomic ID with a
\emph{Semantic ID (SID)}: a short tuple of discrete codes $(c_0, c_1, \dots, c_{L-1})$ produced by a
residual-quantized autoencoder over a content embedding of the item \citep{tiger}. A sequence model
is then trained to \emph{generate} the SID of the next relevant item token-by-token, so retrieval
becomes autoregressive decoding over a small vocabulary rather than a dot-product search over
millions of items. The appeal is threefold: the SID is compact and shareable; semantically similar
items share SID \emph{prefixes} (coarse-to-fine), aiding long-tail generalization; and a large
language model (LLM) can consume and produce SIDs as ordinary tokens, unifying reasoning and
retrieval.

Autoregressive SID generation predicts $c_0$ first, then $c_1 \mid c_0$, and so on. A wrong $c_0$
sends decoding down the wrong branch of the SID tree and corrupts every subsequent code; the coarse
level therefore dominates end-to-end quality. Yet in the standard reconstruction-only RQ-VAE, $c_0$
is an \emph{arbitrary geometric split} of embedding space with no semantic identity. That space is
produced by a multimodal encoder, so the split can turn on signal that never reaches the
retriever: from the user's history alone, the model must still infer which region of that latent
space an item falls into---a hard, opaque prediction.

The deeper problem is a \emph{double} misalignment. The tokenizer is trained \emph{independently} to
reconstruct an item embedding, so its codes are
aligned with neither the LLM that must generate them (they enter the vocabulary as opaque tokens) nor
the end task they must serve. Prior SID systems bridge this gap from two sides:
\emph{consumption-side}, making opaque codes legible to the LLM via alignment corpora, reasoning/RL,
or dedicated per-token encoders \citep{sidreasoner,prefixmem}; and \emph{tokenizer-side}, learning to
make codes task- or structure-aware---back-propagating the task objective into the tokenizer
\citep{diger,bloger} or supervising each level to predict a learned tag hierarchy \citep{hidvae}.
Both \emph{recover or align} meaning after the fact; and because that meaning is content-derived, it
need not be the meaning the task requires.

We therefore make the coarse codes \emph{mean something}. Given any predefined categorical attribute of an
item that we can also place in the model's text prompt---its targeting country, its content
category, its language---we force the coarse quantization index to the attribute's label
instead of the nearest codebook entry, while letting the codebook vectors keep training under the
reconstruction loss (Section~\ref{sec:method-guided}). The guided level is then 100\%
accurate by construction, human-interpretable, and---because the attribute also appears in the
prompt---\emph{text-grounded}: the model reads ``United States'' and emits the country code, turning
coarse-level prediction from inference into lookup. A trie-merge construction maps any
high-cardinality or set-valued attribute onto the fixed code budget while keeping merged buckets
coherent, and we deliberately choose an attribute the task depends on (e.g., country eligibility) so
the coarse level is \emph{task-relevant}, not merely interpretable. The residual levels remain free,
standard RQ-VAE codes that refine content within the guided partition.

\paragraph{Contributions.}
We contribute two things. The first is \textbf{Guided SID} itself: a way to make the coarse levels of
a Semantic ID meaningful \emph{by construction}, replacing nearest-centroid search at the guided
level with deterministic supervised index assignment of a predefined, text-grounded, task-relevant
attribute, while leaving the codebooks learnable and every residual level untouched. The method is
not tied to the attribute, the depth, or the level ordering we happen to use. The second is the
\textbf{trie-merge codebook construction} that makes the first practical: real attributes are
high-cardinality, set-valued and unbalanced, and trie-merge maps any such attribute onto a fixed code
budget by canonicalizing values into prefix paths and repeatedly merging the two smallest siblings,
so that merged buckets stay semantically coherent instead of arbitrary (Section~\ref{sec:method}).

Our experiments then support these two claims. Guiding costs nothing intrinsically---reconstruction
and collision match or beat the vanilla baseline despite spending level~0 on a
non-reconstruction signal (Section~\ref{sec:intrinsic})---and it pays off end to end: in a matched
A/B differing \emph{only} in the SID encoding, the guided retriever leads at every list length
($1.36\times$ recall@1, $1.39\times$ recall@10, MRR $0.0355$ against $0.0260$, against the full
654--741M catalog) and predicts the pre-defined attribute $4.2\times$ more often
(Section~\ref{sec:retrieval}). A prepended-token arm
that leaves the content codes untouched recovers almost none of that gain, which places the effect
in the restructured coarse code rather than in conditioning. Guiding \emph{only} the coarse level also lifts the \emph{free residual} levels, which no scheme
supervises, in all $24$ position-$k$ cells we measure---modestly, at $1.4$--$1.5\times$, rather than
the order of magnitude a smaller candidate set appeared to show (Section~\ref{sec:mechanism}).

\section{Related Work}

\paragraph{Generative retrieval and SID tokenization.}
TIGER \citep{tiger} established the RQ-VAE $\rightarrow$ autoregressive-generation paradigm we build
on. RQ-VAE and RQ-KMeans are the dominant quantizers; both learn \emph{all} levels without any
predefined per-level meaning. OneRec's entropy analysis \citep{onerec} shows SID hierarchies are
naturally coarse-to-fine (per-level entropy falls sharply with depth), motivating placing
interpretable, high-value structure at the coarse level---where Guided SID intervenes.

\paragraph{Improving the quantizer.}
A line of work adds \emph{supervision inside the quantizer}. PLUM/SIDv2 \citep{plum} adds a
co-occurrence contrastive loss on quantized codes (+4.9\% uniqueness / +1.8\% recall); this is the
closest work on ``supervise the quantizer,'' but the signal is \emph{self-supervised behavioral} and
the hierarchy still \emph{emerges} rather than being \emph{imposed}. ADC-SID
\citep{adcsid} and QuaSID \citep{quasid} add behavioral-contrastive or collision-repulsion
terms. ReSID/GAOQ \citep{resid} targets our \emph{goal}---predictable, prefix-unambiguous
coarse levels for autoregressive decoding---but via \emph{unsupervised} global orthogonal alignment;
its codes carry no human meaning and cannot drive attribute-level constrained decoding.
REG4Rec \citep{reg4rec} removes the hierarchy entirely (parallel, order-invariant), the
opposite design axis. Balance-oriented methods (GPR \citep{gpr}, OneRec balanced
K-Means \citep{onerec}) use deterministic/balanced assignment purely for codebook \emph{balance}.

\paragraph{Dedicating a level to a specific signal.}
The single closest prior art is GR4AD/UA-SID \citep{gr4ad}, which dedicates the
\emph{final} SID level to a \emph{non-learnable hash} of non-semantic business features to reduce
collisions (85\%$\rightarrow$18\%). Guided SID differs on all three axes: \emph{which} level
(first/coarse vs.\ final), \emph{how} (a learnable supervised codebook vs.\ a fixed hash), and
\emph{why} (LLM predictability, interpretability, constrained decoding vs.\ collision reduction).
FedMM \citep{fedmm} assigns distinct roles to distinct codebook layers
(federated-shared vs.\ market-local)---the same ``structured layers'' spirit---but the role is set by
federation/privacy, not supervised content attributes.

\paragraph{Making SID meaning \emph{legible} vs.\ \emph{assigned} (our position).}
Closest to us, HiD-VAE \citep{hidvae} supervises \emph{every} RQ-VAE level to align with an
LLM-generated multi-level tag hierarchy, giving interpretable per-level codes; other tokenizer-side
methods make codes task-aware by back-propagating the recommendation objective into the tokenizer
(DIGER \citep{diger}; bi-level optimization, BLOGER \citep{bloger}) or by
collaborative/diversity regularization (LETTER \citep{letter}). On the consumption side,
SIDReasoner and LC-Rec \citep{sidreasoner,lcrec} teach the LLM the meaning of
\emph{whole-item} SIDs via alignment/reasoning data, and PrefixMem \citep{prefixmem} adds a
prefix-conditioned encoder precisely because ``a SID token's meaning depends on its prefix.'' All of
these \emph{recover, align, or explain} a meaning that the (largely unsupervised) codes happen to
carry---and that content-derived meaning may not be task-useful. Guided SID instead \emph{assigns}
the coarse levels' meaning by construction: \emph{deterministic} (an exact index assignment, unlike
HiD-VAE's learned, soft tag-alignment), \emph{externally defined} (a real attribute the task depends
on, rather than content-derived tags), text-grounded so the coarse token becomes an autoregressive
lookup, and directly usable for per-attribute constrained decoding and eligibility filtering. We are
also the first to show, end-to-end, that guiding only the coarse levels makes the \emph{free
residual} levels more learnable.

\paragraph{SID consumption / LLM reasoning.}
SIDReasoner \citep{sidreasoner} grounds opaque RQ-VAE SID tokens for the LLM via an enriched
SID--language alignment corpus plus outcome-driven RL (rewarding correct coarse prefixes)---SID
tokens must be made \emph{legible} to the LLM---but that meaning is \emph{learned}, not assigned, and
attached to the \emph{whole item} SID rather than to individual levels. OneRanker
\citep{oneranker} shows pure SID generation is target-agnostic. Guided SID makes the coarse tokens
legible \emph{by construction}---a level \emph{is} a predefined attribute, so shared-prefix
commonality and per-position semantics are definitional rather than something the LLM must be
taught---and gives partial coarse targeting (e.g., country eligibility) for
free. PinRec \citep{pinrec} is the SID-skeptic foil (SIDs collapse at very large scale); Guided
SID's forced coarse partition is a \emph{mitigation}---it guarantees at least attribute-cardinality
distinct coarse buckets. RecJPQ \citep{recjpq} uses joint product quantization for
embedding-table compression, learned end-to-end from the rec loss with no predefined per-level
attribute---an orthogonal (memory) motivation.

\section{Method}
\label{sec:method}

\subsection{Problem formulation}
An item $x$ has a content embedding $z = f(x) \in \mathbb{R}^{d}$ from a \emph{frozen} multimodal
encoder. An \emph{RQ-VAE tokenizer} turns $z$ into an SID $(c_0,\dots,c_{L-1})$,
$c_\ell \in \{0,\dots,K-1\}$. Its own encoder $E$ first projects the embedding into a low-dimensional
quantization latent $h = E(z) \in \mathbb{R}^{d'}$ ($d' \ll d$). That latent is quantized
\emph{residually}: with $r_0 = h$, level $\ell$ selects
$c_\ell = \arg\min_k \lVert r_\ell - C_\ell[k] \rVert^2$ from its own codebook $C_\ell$ and passes on
$r_{\ell+1} = r_\ell - C_\ell[c_\ell]$, so each level encodes only what earlier levels left over. The
selected codewords sum to the quantized latent $\hat h = \sum_\ell C_\ell[c_\ell]$, which the decoder
$D$ maps back to the embedding space, $\hat z = D(\hat h)$. The codebooks $\{C_\ell\}$ and the pair
$(E,D)$ are trained with
$\mathcal{L} = \lVert z - \hat z \rVert^2 + \beta\,\mathcal{L}_{\text{commit}} +
\mathcal{L}_{\text{codebook}}$; the straight-through estimator (STE) passes gradients through the
discrete selection. A \emph{generative retriever} (an LLM) is then
trained to output an item's SID from context, and retrieval is constrained autoregressive decoding
over the SID vocabulary. We use $L=6$, $K=256$ throughout.

\subsection{Guided assignment}
\label{sec:method-guided}
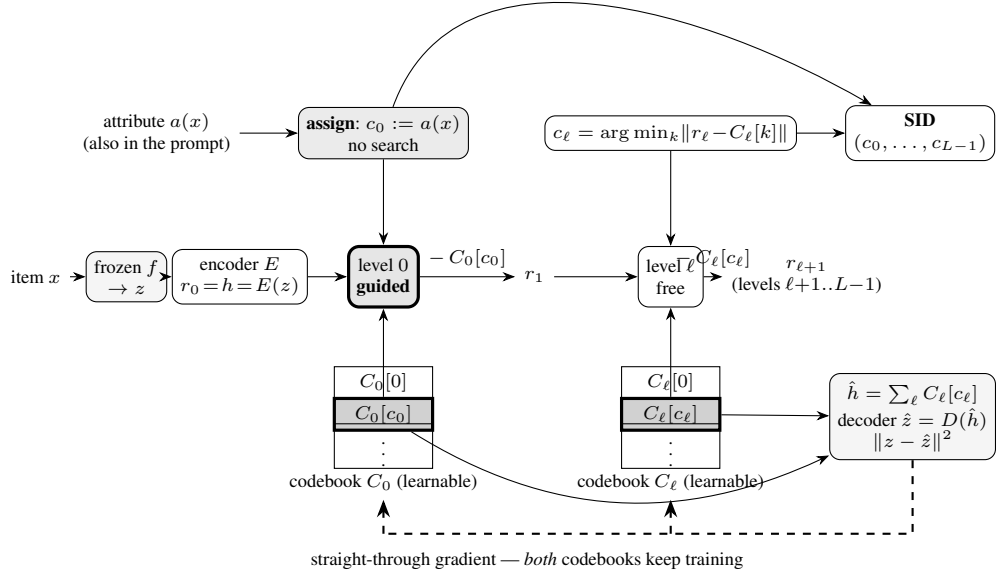
\begin{figure}[t]
\centering
\begin{tikzpicture}[
  >=Stealth, font=\small,
  cb/.style   ={draw, minimum width=13mm, minimum height=4.2mm, inner sep=1pt, font=\scriptsize},
  hit/.style  ={cb, fill=black!18, very thick},
  op/.style   ={draw, rounded corners, align=center, font=\scriptsize, inner sep=3pt},
  note/.style ={font=\scriptsize, align=center}
]
%%% ---------- main residual chain (RQ-VAE), y = 1.7 ----------
\node[note] (x)  at (-0.3,1.7) {item $x$};
\node[op, fill=black!4] (enc) at (0.9,1.7) {frozen $f$\\ \scriptsize $\rightarrow z$};
\node[op] (E) at (2.4,1.7) {encoder $E$\\ \scriptsize $r_0\!=\!h\!=\!E(z)$};
\node[op, fill=black!10, very thick, minimum height=8mm] (q0) at (4.3,1.7) {level $0$\\ \textbf{guided}};
\node[note] (r1) at (6.3,1.7) {$r_1$};
\node[op, minimum height=8mm] (q1) at (8.1,1.7) {level $\ell$\\ free};
\node[note] (r2) at (9.9,1.7) {$r_{\ell+1}$\\ \scriptsize (levels $\ell{+}1..L{-}1$)};
\draw[->] (x) -- (enc);  \draw[->] (enc) -- (E);  \draw[->] (E) -- (q0);
\draw[->] (q0) -- node[above, note] {$-\,C_0[c_0]$} (r1);
\draw[->] (r1) -- (q1);
\draw[->] (q1) -- node[above, note] {$-\,C_\ell[c_\ell]$} (r2);

%%% ---------- index selection (the ONLY difference), above ----------
\node[note] (attr) at (1.3,3.6) {attribute $a(x)$\\ (also in the prompt)};
\node[op, fill=black!8] (s0) at (4.3,3.6) {\textbf{assign}: $c_0 := a(x)$\\ \scriptsize no search};
\node[op] (s1) at (8.1,3.6) {$c_\ell=\arg\min_k\lVert r_\ell\!-\!C_\ell[k]\rVert$};
\draw[->] (attr) -- (s0);
\draw[->] (s0) -- (q0);
\draw[->] (s1) -- (q1);

%%% ---------- per-level codebooks, below ----------
\node[cb]  (a1) at (4.3,0.30) {$C_0[0]$};
\node[hit] (a2) at (4.3,-0.12) {$C_0[c_0]$};
\node[cb]  (a3) at (4.3,-0.54) {$\vdots$};
\node[note] at (4.3,-1.02) {codebook $C_0$ (learnable)};
\draw[->] (a2) -- (q0);

\node[cb]  (b1) at (8.1,0.30) {$C_\ell[0]$};
\node[hit] (b2) at (8.1,-0.12) {$C_\ell[c_\ell]$};
\node[cb]  (b3) at (8.1,-0.54) {$\vdots$};
\node[note] at (8.1,-1.02) {codebook $C_\ell$ (learnable)};
\draw[->] (b2) -- (q1);

%%% ---------- outputs ----------
\node[op] (sid) at (11.4,3.6) {\textbf{SID}\\ $(c_0,\dots,c_{L-1})$};
\draw[->] (s1) -- (sid);
\draw[->] (s0) to[out=70,in=150] (sid);
\node[op, fill=black!4] (rec) at (11.3,-0.15) {$\hat h=\sum_\ell C_\ell[c_\ell]$\\ decoder $\hat z = D(\hat h)$\\ $\lVert z-\hat z\rVert^2$};
\draw[->] (b2) -- (rec);
\draw[->] (a2) to[out=-32,in=205] (rec);

%%% ---------- STE gradient back to BOTH codebooks ----------
\draw[dashed, thick] (rec.south) -- (11.3,-1.7) -- (4.3,-1.7);
\draw[->, dashed, thick] (8.1,-1.7) -- (8.1,-1.25);
\draw[->, dashed, thick] (4.3,-1.7) -- (4.3,-1.25);
\node[note] at (6.2,-2.05) {straight-through gradient --- \emph{both} codebooks keep training};
\end{tikzpicture}
\caption{\textbf{Guided SID inside the RQ-VAE: what changes and what does not.} The horizontal chain
is standard residual quantization: the frozen content embedding $z$ is projected by the tokenizer's
encoder $E$ to the quantization latent $h$, and each level then subtracts its chosen codeword and
passes the residual on, so level $\ell$ quantizes only what earlier levels left over and every level
owns a separate learnable codebook. Reconstruction closes the loop on the right---the selected
codewords sum to $\hat h$, the decoder $D$ maps it back to $\hat z$, and $\lVert z-\hat z\rVert^2$
drives training---while the selected indices, read off the selectors, are the SID. The \emph{sole}
modification is the index selection drawn above each level: a guided level \emph{assigns} the index
from the attribute label ($c_0:=a(x)$, skipping the nearest-neighbour search), a free level searches
as usual. Nothing else changes; in particular the gradient still reaches \emph{both} codebooks
through the straight-through estimator, so the guided codebook keeps training toward the
reconstruction-optimal representative of each attribute value instead of being frozen to a centroid. This
design can also be interpreted as forcing the autoencoder to learn what the samples sharing an
assigned attribute value have in common, in the codebook entry at the guided position. Because guidance changes the
residual handed to level~1, it also reshapes what the \emph{free} levels must learn---the mechanism
examined in Section~\ref{sec:mechanism}.}
\label{fig:method}
\end{figure}

For a chosen set of \emph{guided levels} $\mathcal{G} \subseteq \{0,\dots,L-1\}$ (in practice the
coarse prefix $\{0\}$ or $\{0,1\}$) and a per-item \emph{attribute label} $a_\ell(x) \in
\{0,\dots,K-1\}$ for each $\ell \in \mathcal{G}$, we override the index selection at guided
levels:
\begin{equation}
c_\ell =
\begin{cases}
a_\ell(x) & \ell \in \mathcal{G} \quad\text{(guided: use the attribute label)}\\[2pt]
\arg\min_k \lVert r_\ell - C_\ell[k]\rVert^2 & \ell \notin \mathcal{G} \quad\text{(free: nearest neighbor)}
\end{cases}
\end{equation}
Everything else is unchanged: we still look up $C_\ell[c_\ell]$, still form $r_{\ell+1} = r_\ell -
C_\ell[c_\ell]$, and the codebook row $C_\ell[a_\ell(x)]$ \textbf{still receives reconstruction
gradients} via the STE. The guided codebook is therefore \emph{learned}, not frozen to attribute
centroids---it converges to the reconstruction-optimal representative of each attribute value,
giving 100\% attribute accuracy at the guided level (the index is forced) and
end-to-end reconstruction optimization. Implementation is a small subclass of the quantizer that
accepts \texttt{guided\_ids} and skips the nearest-neighbor search at guided levels; no other
training machinery changes; Figure~\ref{fig:method} illustrates the scheme. The assignment is hard by
design: it guarantees the attribute encoding exactly and adds no hyperparameters.

\subsection{Codebook construction: mapping an attribute to \texorpdfstring{$K$}{K} codes}
\label{sec:codebook}
Assigning a coarse level to an attribute is only feasible if the attribute fits the code budget,
which it rarely does: targeting country spans 200+ codes and ${\sim}481$K co-targeting
\emph{combinations} (an item may target a \emph{set} of countries), and a content taxonomy has
${\sim}6$K leaf paths. A naive ``top-$K$ values $+$ one catch-all'' map is not an option---it dumps
the long tail into a single degenerate code, manufacturing exactly the collapsed codebook that makes
the assigned meaning useless (Section~\ref{sec:limitations}). We instead compress \emph{any} cardinality
to exactly $K$ with a trie-merge.

\textbf{Canonicalize, then merge.} Each attribute value becomes a path in a trie whose siblings share
a prefix: a set-valued attribute is sorted into a canonical sequence (for country, the targeted codes
ordered by global popularity, so related combinations share long prefixes), and a taxonomy attribute
uses its category path directly. Every node stores its own leaf count (items with that exact value)
and its subtree count (all descendants). We then merge \emph{bottom-up}: repeatedly take the parent
whose smallest child holds the fewest items and merge its two smallest children into one leaf,
stopping at $K$ leaves (Figure~\ref{fig:trie}). A priority queue makes this $O(N\log N)$; pseudocode
and statistics are in Appendix~\ref{app:codebook}.

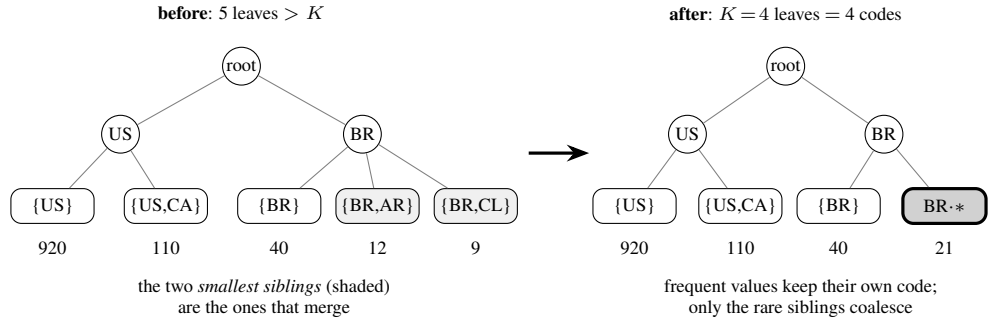
\begin{figure}[t]
\centering
\begin{tikzpicture}[
  >=Stealth, font=\small,
  nd/.style   ={draw, circle, minimum size=5mm, inner sep=0pt, font=\scriptsize},
  lf/.style   ={draw, rounded corners, minimum width=11mm, minimum height=4.6mm, inner sep=1pt, font=\scriptsize},
  small/.style={lf, fill=black!6},
  mrg/.style  ={lf, fill=black!18, very thick},
  note/.style ={font=\scriptsize, align=center},
  ed/.style   ={-, gray}
]
%%%%%%%%%%%%%%%% BEFORE %%%%%%%%%%%%%%%%
\node[note] at (2.5,3.15) {\textbf{before}: 5 leaves $>K$};
\node[nd] (r) at (2.5,2.45) {\scriptsize root};
\node[nd] (us) at (0.9,1.55) {US};
\node[nd] (br) at (4.1,1.55) {BR};
\draw[ed] (r) -- (us); \draw[ed] (r) -- (br);

\node[lf]    (a1) at (0.0,0.6) {\{US\}};
\node[lf]    (a2) at (1.5,0.6) {\{US,CA\}};
\node[lf]    (a3) at (3.0,0.6) {\{BR\}};
\node[small] (a4) at (4.3,0.6) {\{BR,AR\}};
\node[small] (a5) at (5.6,0.6) {\{BR,CL\}};
\draw[ed] (us) -- (a1); \draw[ed] (us) -- (a2);
\draw[ed] (br) -- (a3); \draw[ed] (br) -- (a4); \draw[ed] (br) -- (a5);
\node[note] at (0.0,0.05) {920};
\node[note] at (1.5,0.05) {110};
\node[note] at (3.0,0.05) {40};
\node[note] at (4.3,0.05) {12};
\node[note] at (5.6,0.05) {9};
\node[note] at (2.8,-0.62) {the two \emph{smallest siblings} (shaded)\\ are the ones that merge};

%%%%%%%%%%%%%%%% arrow %%%%%%%%%%%%%%%%
\draw[->, very thick] (6.3,1.3) -- (7.1,1.3);

%%%%%%%%%%%%%%%% AFTER %%%%%%%%%%%%%%%%
\node[note] at (9.7,3.15) {\textbf{after}: $K\!=\!4$ leaves $=$ 4 codes};
\node[nd] (r2) at (9.7,2.45) {\scriptsize root};
\node[nd] (us2) at (8.4,1.55) {US};
\node[nd] (br2) at (11.0,1.55) {BR};
\draw[ed] (r2) -- (us2); \draw[ed] (r2) -- (br2);

\node[lf]  (b1) at (7.7,0.6) {\{US\}};
\node[lf]  (b2) at (9.1,0.6) {\{US,CA\}};
\node[lf]  (b3) at (10.4,0.6) {\{BR\}};
\node[mrg] (b4) at (11.8,0.6) {BR$\cdot\ast$};
\draw[ed] (us2) -- (b1); \draw[ed] (us2) -- (b2);
\draw[ed] (br2) -- (b3); \draw[ed] (br2) -- (b4);
\node[note] at (7.7,0.05) {920};
\node[note] at (9.1,0.05) {110};
\node[note] at (10.4,0.05) {40};
\node[note] at (11.8,0.05) {21};
\node[note] at (9.9,-0.62) {frequent values keep their own code;\\ only the rare siblings coalesce};
\end{tikzpicture}
\caption{\textbf{Trie-merge: any attribute onto a fixed budget of $K$ codes.} Values are canonicalized
into paths so that \emph{siblings share a prefix} (here, co-targeted country sets), and each step
merges the two smallest siblings---necessarily the two most similar values. Frequent values
(\{US\}, \{BR\}) are never the smallest sibling and keep dedicated codes, while the rare relatives
\{BR,AR\} and \{BR,CL\} coalesce into one still-coherent code (BR$\cdot\ast$, ``Brazil plus a small
neighbour''); a top-$K$ cut would instead sweep every rare value into a single meaningless catch-all.
Merging is always \emph{within} a parent, so a prefix keeps as many codes as it has surviving
children; in the limit where all of a parent's children merge, that subtree collapses to one code
identified by the prefix itself. Item counts are illustrative.}
\label{fig:trie}
\end{figure}

\textbf{Why a trie.} Because siblings share a prefix, the algorithm only ever merges the \emph{most
similar} values (neighboring countries, sibling categories)---no $O(N^2)$ pairwise-similarity search
and no arbitrary catch-all. Large values are never the smallest sibling, so they keep their own code;
small values fold into their closest relatives. The result is deterministic, reproducible, and
balanced \emph{where the data allows}.

\begin{figure}[t]
\centering
\begin{tikzpicture}[
  >=Stealth, font=\small,
  nd/.style   ={draw, circle, minimum size=4.6mm, inner sep=0pt, font=\tiny},
  lf/.style   ={draw, rounded corners, minimum width=9mm, minimum height=4.4mm, inner sep=1pt, font=\tiny},
  hdr/.style  ={font=\scriptsize\bfseries, align=center},
  note/.style ={font=\scriptsize, align=center},
  ed/.style   ={-, gray}
]
%%%%% (a) taxonomy
\node[hdr] at (1.6,3.5) {(a) taxonomic};
\node[note] at (1.6,3.05) {category path};
\node[nd] (t0) at (1.6,2.4) {};
\node[nd] (t1) at (0.8,1.6) {};
\node[nd] (t2) at (2.4,1.6) {};
\node[lf] (t3) at (0.3,0.75) {};
\node[lf] (t4) at (1.3,0.75) {};
\node[lf] (t5) at (2.4,0.75) {};
\draw[ed] (t0)--(t1); \draw[ed] (t0)--(t2);
\draw[ed] (t1)--(t3); \draw[ed] (t1)--(t4); \draw[ed] (t2)--(t5);
\node[note] at (1.6,0.15) {path is the trie};

%%%%% (b) set-valued
\node[hdr] at (5.6,3.5) {(b) set-valued};
\node[note] at (5.6,3.05) {sort by popularity};
\node[note] at (5.6,2.45) {$\{$CA,US$\}\!\rightarrow\!$US$\cdot$CA};
\node[nd] (s0) at (5.6,1.85) {};
\node[nd] (s1) at (4.9,1.15) {};
\node[nd] (s2) at (6.3,1.15) {};
\node[lf] (s3) at (4.5,0.5) {};
\node[lf] (s4) at (5.5,0.5) {};
\node[lf] (s5) at (6.5,0.5) {};
\draw[ed] (s0)--(s1); \draw[ed] (s0)--(s2);
\draw[ed] (s1)--(s3); \draw[ed] (s1)--(s4); \draw[ed] (s2)--(s5);
\node[note] at (5.6,-0.1) {sorted set is the trie};

%%%%% (c) continuous
\node[hdr] at (9.9,3.5) {(c) continuous};
\node[note] at (9.9,3.05) {recursive quantile split};
\draw[|-|, thick] (8.6,2.5) -- (11.2,2.5);
\draw (9.9,2.65) -- (9.9,2.35);
\draw (9.25,2.62) -- (9.25,2.38);
\draw (10.55,2.62) -- (10.55,2.38);
\node[nd] (u0) at (9.9,1.75) {};
\node[nd] (u1) at (9.25,1.1) {};
\node[nd] (u2) at (10.55,1.1) {};
\node[lf] (u3) at (8.85,0.5) {};
\node[lf] (u4) at (9.85,0.5) {};
\node[lf] (u5) at (10.85,0.5) {};
\draw[ed] (u0)--(u1); \draw[ed] (u0)--(u2);
\draw[ed] (u1)--(u3); \draw[ed] (u1)--(u4); \draw[ed] (u2)--(u5);
\node[note] at (9.9,-0.1) {siblings $=$ adjacent ranges};

%%%%% shared merge
\node[draw, rounded corners, very thick, minimum width=112mm, minimum height=8mm,
      align=center, font=\scriptsize] (merge) at (5.6,-1.1)
  {\textbf{same bottom-up merge}: repeatedly combine the two smallest siblings until $K$ leaves remain
   $\;\Rightarrow\;$ one code per leaf};
\draw[->] (1.6,-0.4) -- (1.6,-0.72);
\draw[->] (5.6,-0.4) -- (5.6,-0.72);
\draw[->] (9.9,-0.4) -- (9.9,-0.72);
\end{tikzpicture}
\caption{\textbf{The construction is attribute-agnostic.} All that is required is a hierarchical
decomposition in which \emph{siblings are similar}: a taxonomic attribute supplies one directly (a);
a set-valued attribute becomes one after sorting into a canonical order (b); and a continuous
attribute becomes one under recursive quantile splitting, where siblings are adjacent ranges (c). The
identical bottom-up sibling merge then compresses any of them to exactly $K$ codes. Representability
is not sufficiency, however: a guided level also needs its value to be recoverable from the prompt,
which is why we surface a binned continuous attribute by a semantic bin name rather than a bin index,
and why we evaluate categorical attributes only.}
\label{fig:general}
\end{figure}
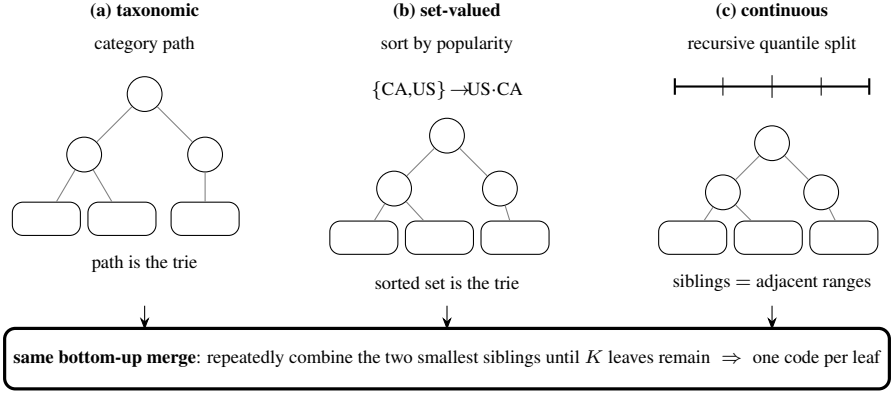

\textbf{Beyond categorical attributes.} The construction requires only that values admit a
hierarchical decomposition in which siblings are similar (Figure~\ref{fig:general}); it is not
restricted to categorical
attributes. A continuous attribute qualifies via recursive quantile splitting, whose split tree makes
siblings \emph{adjacent ranges}, so merging the two smallest siblings again merges the two most
similar values. Two caveats temper this generality. First, for a plain scalar the construction adds
little: quantile binning to exactly $K$ is already balanced by definition, and trie-merge earns its
keep on high-cardinality categorical, set-valued, or taxonomic attributes where a top-$K$ cut would
destroy coherence. Second---and more importantly---being \emph{representable} as $K$ codes does not
make an attribute a good \emph{guide}: a guided level pays off only when its value is recoverable
from the prompt, and asking a model to map a numeral (``\$47.32'') to an
arbitrary bin index is precisely the kind of digit-to-index association that autoregressive models
learn poorly. A binned continuous attribute should therefore be surfaced in the prompt by a
\emph{semantic bin name} (``price tier: premium'') rather than a bin index. We evaluate categorical
attributes only.

\subsection{Feature selection}
The guiding attribute should be as intrinsic to the item and as available at inference time as
possible. Only an attribute that actually reflects the item's content can build a meaningful
codebook, and a guiding attribute helps the model only when it can be hinted at inference
time---otherwise the coarse code is still guessed. Targeting information such as country or language
is normally a good guiding attribute in an ads ranking system, since advertisers specialize their
ads according to the audiences they target. Category may also help when users can filter by category
in a search system.

Beyond these two, an attribute must be text-groundable so the model can connect prompt to code,
well covered, stable over time, and codebook-viable---high-cardinality or set-valued attributes have
to survive the merge of Section~\ref{sec:codebook} with coherent buckets. Our experiments guide on
targeting country, and we return to what that choice costs---a head-weighted codebook whose
buckets track content indirectly---in Section~\ref{sec:limitations}.

\section{Experimental Setup}
\label{sec:setup}

Our experiments compare three SID encodings: the vanilla (reconstruction-only) SID, the proposed
guided SID, and a third \emph{prepended} SID that leaves the content codes untouched and instead
emits the attribute as an extra leading token (Appendix~\ref{app:prepend}). The two contrasts
isolate different things. Prepended versus vanilla measures what the attribute contributes as
\emph{extra information} the model is conditioned on---the ad targeting country, in our
experiment---while the content codes stay identical. Guided versus prepended measures what is gained
by folding that attribute into RQ-VAE training itself rather than carrying it alongside.

\paragraph{Data and embeddings.}
All experiments use a large industrial ads corpus: the tokenizers are minted over $\sim$4.4B ads,
each with a 512-dim frozen multimodal content embedding and a structured text description that
already contains the targeting attributes. One caveat is that these production ads are less curated than a
research corpus: missing or misaligned ad details and absent actions degrade the final result.

\paragraph{Tokenizers.} \textbf{Vanilla (baseline)}---a production reconstruction-only RQ-VAE,
$256\times6$, all levels free; \textbf{Guided}---identical architecture and training, but level~0
forced to the targeting-country code (256-way, trie-merged), levels~1--5 free. In the end-to-end A/B
the two SIDs are \emph{minted together from the same embedding dump} and joined 1:1, so coverage is
${\sim}$identical (${\sim}100\%$) and the only variable is the encoding.

\paragraph{Generative retriever (CPT $\rightarrow$ SFT).}
The retriever is an LLM (Qwen3-4B) whose vocabulary includes the $256\times6$ SID tokens. Training
has two stages, identical across arms except for the SID values: \emph{continual pretraining (CPT)},
20{,}000 steps, seq-len 8192, 70\% ads / 30\% public text, AdamW, lr $6\mathrm{e}{-}6 \rightarrow
3\mathrm{e}{-}7$ cosine (400 warmup); and \emph{supervised finetuning (SFT)}, 6{,}000 steps on
user-history $\rightarrow$ next-ad-SID, matched across arms.

\paragraph{Metrics and protocol.}
\emph{Intrinsic}: collision rate (fraction of ads without a unique full SID; lower better),
reconstruction distance (lower better), and per-prefix cluster precision / separation recall against
ground-truth attributes. \emph{End-to-end}: greedy constrained decoding over a trie spanning
a shared $10{,}000$-ad catalog identical across arms ($N=10{,}000$ per arm, avg history 19.3 ads);
we report recall@1
(exact 6/6 match), country\_match@1, and per-codebook top-1 accuracy, each compared directly
between arms.

\section{Results}
\label{sec:results}
\label{sec:intrinsic}
\label{sec:retrieval}

Forcing level~0 to encode country does not degrade the tokenizer---it improves it
(Table~\ref{tab:intrinsic}): reconstruction distance is 27\% lower than the
vanilla baseline despite level~0 being spent on a non-reconstruction signal. Cluster
precision against content-category ground truth also grows cleanly with prefix depth, from $0.414$
at the country-only prefix to $0.956$ at the full SID, confirming a coarse-to-fine hierarchy in
which the free residual levels recover the category resolution level~0 spends on country
(Appendix~\ref{app:intrinsic}).

\begin{table}[t]
\centering
\caption{Intrinsic tokenizer quality (country-guided at L0, $256\times6$). Guiding the coarse level improves reconstruction.}
\label{tab:intrinsic}
\begin{tabular}{lccc}
\toprule
Metric & \textbf{Guided} & Vanilla baseline & LLM-CLS \\
\midrule
Collision@1 (lower better) & \textbf{88.4\%} & 91.9\% & 88.3\% \\
Avg reconstruction distance & \textbf{0.085} & 0.117 & 0.110 \\
P95 reconstruction distance & \textbf{0.124} & 0.173 & 0.152 \\
\bottomrule
\end{tabular}
\end{table}

With the \emph{only} difference being the SID encoding, the guided retriever wins decisively
(Table~\ref{tab:retrieval}): it leads at every list length and carries the higher mean reciprocal
rank, and $18\%$ of its recommendations land in the correct country against $4\%$ for the baseline.
This is measured against the \emph{full} production catalog, so no candidate-set sampling enters the
comparison.

\begin{table}[t]
\centering
\caption{Matched end-to-end retrieval A/B ($N=10{,}000$/arm, sampled
decoding at $T{=}1.0$ with ten draws per record) against the \textbf{full production catalog of
654--741M identifiers}. Only the SID encoding differs; the serving-known attribute is supplied in
the prompt to every arm, including the baseline. Rates are over all records, so the denominators
are identical across arms.}
\label{tab:retrieval}
\begin{tabular}{lcccc}
\toprule
Metric & \textbf{Guided} & Vanilla baseline & Prepended & Guided/base \\
\midrule
recall@1  & \textbf{0.0166} & 0.0122 & 0.0076 & \textbf{$1.36\times$} \\
recall@3  & \textbf{0.0417} & 0.0318 & 0.0219 & \textbf{$1.31\times$} \\
recall@5  & \textbf{0.0614} & 0.0441 & 0.0311 & \textbf{$1.39\times$} \\
recall@10 & \textbf{0.0890} & 0.0639 & 0.0510 & \textbf{$1.39\times$} \\
MRR       & \textbf{0.0355} & 0.0260 & 0.0184 & \textbf{$1.37\times$} \\
pre-defined attribute match@1 & \textbf{0.1770} & 0.0422 & 0.1618 & \textbf{$4.19\times$} \\
\bottomrule
\end{tabular}
\end{table}

\section{Discussion and Limitations}
\label{sec:limitations}
\label{sec:mechanism}

Retrieval is what the encoding is for, so we measure it directly: recall@$k$ over a sampled
candidate list, for all three encodings under one protocol (Table~\ref{tab:retrieval}). Guiding leads
at every list length---$1.66$ against the vanilla baseline's $1.22\%$ at $k{=}1$, rising to $8.90$
against $6.39\%$ at $k{=}10$---and carries the higher mean reciprocal rank, $0.0355$ against
$0.0260$. A third arm that \emph{prepends} the same attribute as an extra token, keeping all six
content codes rather than spending one of them, reaches only $0.76\%$ at $k{=}1$ with an MRR of
$0.0184$: conditioning on the attribute does not recover the gain that comes from restructuring the
coarse code itself.

\begin{figure}[t]
\centering
\begin{tikzpicture}
\begin{axis}[
  width=0.97\linewidth, height=7.4cm,
  % Log y. Values span 0.76%--65.4%, nearly two orders of magnitude; on a linear axis the
  % whole residual region -- what this figure exists to compare -- collapses to stubs.
  ymode=log, log origin=infty,
  enlarge x limits=0.07, ymin=0.5, ymax=600,
  ytick={1,3,10,30,60}, yticklabels={1,3,10,30,60},
  ylabel={prefix recall@$k$ (\%), log scale}, xlabel={SID position},
  symbolic x coords={attr,p0,c1,c2,c3,c4,c5}, xtick={attr,p0,c1,c2,c3,c4,c5},
  xticklabels={attr,pos~0,$c_1$,$c_2$,$c_3$,$c_4$,$c_5$},
  x tick label style={font=\tiny, yshift=-9pt},
  ymajorgrids, grid style={gray!25},
  clip=false,   % the k row and `k =' sit below ymin, outside the axis box
  area legend,
  legend style={at={(0.50,0.99)}, anchor=north, font=\tiny, draw=none, legend columns=3},
  legend cell align=left, font=\scriptsize]
% -------------------------------------------------------------------------------------
% Each position reserves four slots, k=1,3,5,10 (shift -10.5/-3.5/+3.5/+10.5pt). Within a
% slot the three arms share the SAME shift and the SAME width, and are emitted TALLEST
% FIRST so the smallest bar is drawn last and sits in front -- every arm stays visible
% whatever the ordering. That is why the bars are emitted per (position, k) rather than as
% three whole series: pgfplots draws a series in one go, so a fixed arm order cannot adapt,
% and prepending overtakes the baseline at c1 while the baseline leads from c2 down.
% Legend entries come from \addlegendimage since every plot carries `forget plot'.
% Value labels are one rotated line per slot, ordered LARGEST FIRST, so reading outward
% from the bar gives the ranking; colour identifies the arm.
% -------------------------------------------------------------------------------------
\addlegendimage{area legend, fill=blue!55, draw=blue!70!black}\addlegendentry{guided}
\addlegendimage{area legend, fill=orange!75, draw=orange!80!black}\addlegendentry{vanilla}
\addlegendimage{area legend, fill=black, draw=black}\addlegendentry{prepended}
\addplot[ybar, bar shift=-10.5pt, bar width=5pt, fill=black, draw=black, forget plot] coordinates {(attr,16.18)};
\addplot[ybar, bar shift=-3.5pt, bar width=5pt, fill=black, draw=black, forget plot] coordinates {(attr,37.56)};
\addplot[ybar, bar shift=3.5pt, bar width=5pt, fill=black, draw=black, forget plot] coordinates {(attr,49.57)};
\addplot[ybar, bar shift=10.5pt, bar width=5pt, fill=black, draw=black, forget plot] coordinates {(attr,62.45)};
\addplot[ybar, bar shift=-10.5pt, bar width=5pt, fill=blue!55, draw=blue!70!black, forget plot] coordinates {(p0,17.7)};
\addplot[ybar, bar shift=-10.5pt, bar width=5pt, fill=orange!75, draw=orange!80!black, forget plot] coordinates {(p0,4.22)};
\addplot[ybar, bar shift=-10.5pt, bar width=5pt, fill=black, draw=black, forget plot] coordinates {(p0,2.91)};
\addplot[ybar, bar shift=-3.5pt, bar width=5pt, fill=blue!55, draw=blue!70!black, forget plot] coordinates {(p0,40.1)};
\addplot[ybar, bar shift=-3.5pt, bar width=5pt, fill=orange!75, draw=orange!80!black, forget plot] coordinates {(p0,11.18)};
\addplot[ybar, bar shift=-3.5pt, bar width=5pt, fill=black, draw=black, forget plot] coordinates {(p0,8.02)};
\addplot[ybar, bar shift=3.5pt, bar width=5pt, fill=blue!55, draw=blue!70!black, forget plot] coordinates {(p0,51.87)};
\addplot[ybar, bar shift=3.5pt, bar width=5pt, fill=orange!75, draw=orange!80!black, forget plot] coordinates {(p0,16.45)};
\addplot[ybar, bar shift=3.5pt, bar width=5pt, fill=black, draw=black, forget plot] coordinates {(p0,11.64)};
\addplot[ybar, bar shift=10.5pt, bar width=5pt, fill=blue!55, draw=blue!70!black, forget plot] coordinates {(p0,65.38)};
\addplot[ybar, bar shift=10.5pt, bar width=5pt, fill=orange!75, draw=orange!80!black, forget plot] coordinates {(p0,25.57)};
\addplot[ybar, bar shift=10.5pt, bar width=5pt, fill=black, draw=black, forget plot] coordinates {(p0,18.32)};
\addplot[ybar, bar shift=-10.5pt, bar width=5pt, fill=blue!55, draw=blue!70!black, forget plot] coordinates {(c1,2.9)};
\addplot[ybar, bar shift=-10.5pt, bar width=5pt, fill=black, draw=black, forget plot] coordinates {(c1,1.97)};
\addplot[ybar, bar shift=-10.5pt, bar width=5pt, fill=orange!75, draw=orange!80!black, forget plot] coordinates {(c1,1.93)};
\addplot[ybar, bar shift=-3.5pt, bar width=5pt, fill=blue!55, draw=blue!70!black, forget plot] coordinates {(c1,7.25)};
\addplot[ybar, bar shift=-3.5pt, bar width=5pt, fill=black, draw=black, forget plot] coordinates {(c1,5.52)};
\addplot[ybar, bar shift=-3.5pt, bar width=5pt, fill=orange!75, draw=orange!80!black, forget plot] coordinates {(c1,4.89)};
\addplot[ybar, bar shift=3.5pt, bar width=5pt, fill=blue!55, draw=blue!70!black, forget plot] coordinates {(c1,10.43)};
\addplot[ybar, bar shift=3.5pt, bar width=5pt, fill=black, draw=black, forget plot] coordinates {(c1,7.88)};
\addplot[ybar, bar shift=3.5pt, bar width=5pt, fill=orange!75, draw=orange!80!black, forget plot] coordinates {(c1,6.97)};
\addplot[ybar, bar shift=10.5pt, bar width=5pt, fill=blue!55, draw=blue!70!black, forget plot] coordinates {(c1,15.99)};
\addplot[ybar, bar shift=10.5pt, bar width=5pt, fill=black, draw=black, forget plot] coordinates {(c1,12.42)};
\addplot[ybar, bar shift=10.5pt, bar width=5pt, fill=orange!75, draw=orange!80!black, forget plot] coordinates {(c1,10.46)};
\addplot[ybar, bar shift=-10.5pt, bar width=5pt, fill=blue!55, draw=blue!70!black, forget plot] coordinates {(c2,2.35)};
\addplot[ybar, bar shift=-10.5pt, bar width=5pt, fill=orange!75, draw=orange!80!black, forget plot] coordinates {(c2,1.3)};
\addplot[ybar, bar shift=-10.5pt, bar width=5pt, fill=black, draw=black, forget plot] coordinates {(c2,1.06)};
\addplot[ybar, bar shift=-3.5pt, bar width=5pt, fill=blue!55, draw=blue!70!black, forget plot] coordinates {(c2,5.78)};
\addplot[ybar, bar shift=-3.5pt, bar width=5pt, fill=orange!75, draw=orange!80!black, forget plot] coordinates {(c2,3.36)};
\addplot[ybar, bar shift=-3.5pt, bar width=5pt, fill=black, draw=black, forget plot] coordinates {(c2,2.91)};
\addplot[ybar, bar shift=3.5pt, bar width=5pt, fill=blue!55, draw=blue!70!black, forget plot] coordinates {(c2,8.29)};
\addplot[ybar, bar shift=3.5pt, bar width=5pt, fill=orange!75, draw=orange!80!black, forget plot] coordinates {(c2,4.67)};
\addplot[ybar, bar shift=3.5pt, bar width=5pt, fill=black, draw=black, forget plot] coordinates {(c2,4.22)};
\addplot[ybar, bar shift=10.5pt, bar width=5pt, fill=blue!55, draw=blue!70!black, forget plot] coordinates {(c2,12.34)};
\addplot[ybar, bar shift=10.5pt, bar width=5pt, fill=orange!75, draw=orange!80!black, forget plot] coordinates {(c2,6.79)};
\addplot[ybar, bar shift=10.5pt, bar width=5pt, fill=black, draw=black, forget plot] coordinates {(c2,6.76)};
\addplot[ybar, bar shift=-10.5pt, bar width=5pt, fill=blue!55, draw=blue!70!black, forget plot] coordinates {(c3,1.98)};
\addplot[ybar, bar shift=-10.5pt, bar width=5pt, fill=orange!75, draw=orange!80!black, forget plot] coordinates {(c3,1.22)};
\addplot[ybar, bar shift=-10.5pt, bar width=5pt, fill=black, draw=black, forget plot] coordinates {(c3,1.01)};
\addplot[ybar, bar shift=-3.5pt, bar width=5pt, fill=blue!55, draw=blue!70!black, forget plot] coordinates {(c3,4.92)};
\addplot[ybar, bar shift=-3.5pt, bar width=5pt, fill=orange!75, draw=orange!80!black, forget plot] coordinates {(c3,3.18)};
\addplot[ybar, bar shift=-3.5pt, bar width=5pt, fill=black, draw=black, forget plot] coordinates {(c3,2.71)};
\addplot[ybar, bar shift=3.5pt, bar width=5pt, fill=blue!55, draw=blue!70!black, forget plot] coordinates {(c3,7.26)};
\addplot[ybar, bar shift=3.5pt, bar width=5pt, fill=orange!75, draw=orange!80!black, forget plot] coordinates {(c3,4.41)};
\addplot[ybar, bar shift=3.5pt, bar width=5pt, fill=black, draw=black, forget plot] coordinates {(c3,3.94)};
\addplot[ybar, bar shift=10.5pt, bar width=5pt, fill=blue!55, draw=blue!70!black, forget plot] coordinates {(c3,10.62)};
\addplot[ybar, bar shift=10.5pt, bar width=5pt, fill=orange!75, draw=orange!80!black, forget plot] coordinates {(c3,6.4)};
\addplot[ybar, bar shift=10.5pt, bar width=5pt, fill=black, draw=black, forget plot] coordinates {(c3,6.31)};
\addplot[ybar, bar shift=-10.5pt, bar width=5pt, fill=blue!55, draw=blue!70!black, forget plot] coordinates {(c4,1.74)};
\addplot[ybar, bar shift=-10.5pt, bar width=5pt, fill=orange!75, draw=orange!80!black, forget plot] coordinates {(c4,1.22)};
\addplot[ybar, bar shift=-10.5pt, bar width=5pt, fill=black, draw=black, forget plot] coordinates {(c4,1.0)};
\addplot[ybar, bar shift=-3.5pt, bar width=5pt, fill=blue!55, draw=blue!70!black, forget plot] coordinates {(c4,4.44)};
\addplot[ybar, bar shift=-3.5pt, bar width=5pt, fill=orange!75, draw=orange!80!black, forget plot] coordinates {(c4,3.18)};
\addplot[ybar, bar shift=-3.5pt, bar width=5pt, fill=black, draw=black, forget plot] coordinates {(c4,2.69)};
\addplot[ybar, bar shift=3.5pt, bar width=5pt, fill=blue!55, draw=blue!70!black, forget plot] coordinates {(c4,6.45)};
\addplot[ybar, bar shift=3.5pt, bar width=5pt, fill=orange!75, draw=orange!80!black, forget plot] coordinates {(c4,4.41)};
\addplot[ybar, bar shift=3.5pt, bar width=5pt, fill=black, draw=black, forget plot] coordinates {(c4,3.91)};
\addplot[ybar, bar shift=10.5pt, bar width=5pt, fill=blue!55, draw=blue!70!black, forget plot] coordinates {(c4,9.41)};
\addplot[ybar, bar shift=10.5pt, bar width=5pt, fill=orange!75, draw=orange!80!black, forget plot] coordinates {(c4,6.39)};
\addplot[ybar, bar shift=10.5pt, bar width=5pt, fill=black, draw=black, forget plot] coordinates {(c4,6.24)};
\addplot[ybar, bar shift=-10.5pt, bar width=5pt, fill=blue!55, draw=blue!70!black, forget plot] coordinates {(c5,1.66)};
\addplot[ybar, bar shift=-10.5pt, bar width=5pt, fill=orange!75, draw=orange!80!black, forget plot] coordinates {(c5,1.22)};
\addplot[ybar, bar shift=-10.5pt, bar width=5pt, fill=black, draw=black, forget plot] coordinates {(c5,0.76)};
\addplot[ybar, bar shift=-3.5pt, bar width=5pt, fill=blue!55, draw=blue!70!black, forget plot] coordinates {(c5,4.17)};
\addplot[ybar, bar shift=-3.5pt, bar width=5pt, fill=orange!75, draw=orange!80!black, forget plot] coordinates {(c5,3.18)};
\addplot[ybar, bar shift=-3.5pt, bar width=5pt, fill=black, draw=black, forget plot] coordinates {(c5,2.19)};
\addplot[ybar, bar shift=3.5pt, bar width=5pt, fill=blue!55, draw=blue!70!black, forget plot] coordinates {(c5,6.14)};
\addplot[ybar, bar shift=3.5pt, bar width=5pt, fill=orange!75, draw=orange!80!black, forget plot] coordinates {(c5,4.41)};
\addplot[ybar, bar shift=3.5pt, bar width=5pt, fill=black, draw=black, forget plot] coordinates {(c5,3.11)};
\addplot[ybar, bar shift=10.5pt, bar width=5pt, fill=blue!55, draw=blue!70!black, forget plot] coordinates {(c5,8.9)};
\addplot[ybar, bar shift=10.5pt, bar width=5pt, fill=orange!75, draw=orange!80!black, forget plot] coordinates {(c5,6.39)};
\addplot[ybar, bar shift=10.5pt, bar width=5pt, fill=black, draw=black, forget plot] coordinates {(c5,5.1)};
% ---- values, largest first ----
\node[rotate=90, anchor=west, font=\tiny, inner sep=1.2pt]
  at ([xshift=-10.5pt]axis cs:attr,16.18) {\textcolor{black}{16.18}};
\node[rotate=90, anchor=west, font=\tiny, inner sep=1.2pt]
  at ([xshift=-3.5pt]axis cs:attr,37.56) {\textcolor{black}{37.56}};
\node[rotate=90, anchor=west, font=\tiny, inner sep=1.2pt]
  at ([xshift=3.5pt]axis cs:attr,49.57) {\textcolor{black}{49.57}};
\node[rotate=90, anchor=west, font=\tiny, inner sep=1.2pt]
  at ([xshift=10.5pt]axis cs:attr,62.45) {\textcolor{black}{62.45}};
\node[rotate=90, anchor=west, font=\tiny, inner sep=1.2pt]
  at ([xshift=-10.5pt]axis cs:p0,17.7) {\textcolor{black}{2.91}/\textcolor{orange!80!black}{4.22}/\textcolor{blue!70!black}{17.70}};
\node[rotate=90, anchor=west, font=\tiny, inner sep=1.2pt]
  at ([xshift=-3.5pt]axis cs:p0,40.1) {\textcolor{black}{8.02}/\textcolor{orange!80!black}{11.18}/\textcolor{blue!70!black}{40.10}};
\node[rotate=90, anchor=west, font=\tiny, inner sep=1.2pt]
  at ([xshift=3.5pt]axis cs:p0,51.87) {\textcolor{black}{11.64}/\textcolor{orange!80!black}{16.45}/\textcolor{blue!70!black}{51.87}};
\node[rotate=90, anchor=west, font=\tiny, inner sep=1.2pt]
  at ([xshift=10.5pt]axis cs:p0,65.38) {\textcolor{black}{18.32}/\textcolor{orange!80!black}{25.57}/\textcolor{blue!70!black}{65.38}};
\node[rotate=90, anchor=west, font=\tiny, inner sep=1.2pt]
  at ([xshift=-10.5pt]axis cs:c1,2.9) {\textcolor{orange!80!black}{1.93}/\textcolor{black}{1.97}/\textcolor{blue!70!black}{2.90}};
\node[rotate=90, anchor=west, font=\tiny, inner sep=1.2pt]
  at ([xshift=-3.5pt]axis cs:c1,7.25) {\textcolor{orange!80!black}{4.89}/\textcolor{black}{5.52}/\textcolor{blue!70!black}{7.25}};
\node[rotate=90, anchor=west, font=\tiny, inner sep=1.2pt]
  at ([xshift=3.5pt]axis cs:c1,10.43) {\textcolor{orange!80!black}{6.97}/\textcolor{black}{7.88}/\textcolor{blue!70!black}{10.43}};
\node[rotate=90, anchor=west, font=\tiny, inner sep=1.2pt]
  at ([xshift=10.5pt]axis cs:c1,15.99) {\textcolor{orange!80!black}{10.46}/\textcolor{black}{12.42}/\textcolor{blue!70!black}{15.99}};
\node[rotate=90, anchor=west, font=\tiny, inner sep=1.2pt]
  at ([xshift=-10.5pt]axis cs:c2,2.35) {\textcolor{black}{1.06}/\textcolor{orange!80!black}{1.30}/\textcolor{blue!70!black}{2.35}};
\node[rotate=90, anchor=west, font=\tiny, inner sep=1.2pt]
  at ([xshift=-3.5pt]axis cs:c2,5.78) {\textcolor{black}{2.91}/\textcolor{orange!80!black}{3.36}/\textcolor{blue!70!black}{5.78}};
\node[rotate=90, anchor=west, font=\tiny, inner sep=1.2pt]
  at ([xshift=3.5pt]axis cs:c2,8.29) {\textcolor{black}{4.22}/\textcolor{orange!80!black}{4.67}/\textcolor{blue!70!black}{8.29}};
\node[rotate=90, anchor=west, font=\tiny, inner sep=1.2pt]
  at ([xshift=10.5pt]axis cs:c2,12.34) {\textcolor{black}{6.76}/\textcolor{orange!80!black}{6.79}/\textcolor{blue!70!black}{12.34}};
\node[rotate=90, anchor=west, font=\tiny, inner sep=1.2pt]
  at ([xshift=-10.5pt]axis cs:c3,1.98) {\textcolor{black}{1.01}/\textcolor{orange!80!black}{1.22}/\textcolor{blue!70!black}{1.98}};
\node[rotate=90, anchor=west, font=\tiny, inner sep=1.2pt]
  at ([xshift=-3.5pt]axis cs:c3,4.92) {\textcolor{black}{2.71}/\textcolor{orange!80!black}{3.18}/\textcolor{blue!70!black}{4.92}};
\node[rotate=90, anchor=west, font=\tiny, inner sep=1.2pt]
  at ([xshift=3.5pt]axis cs:c3,7.26) {\textcolor{black}{3.94}/\textcolor{orange!80!black}{4.41}/\textcolor{blue!70!black}{7.26}};
\node[rotate=90, anchor=west, font=\tiny, inner sep=1.2pt]
  at ([xshift=10.5pt]axis cs:c3,10.62) {\textcolor{black}{6.31}/\textcolor{orange!80!black}{6.40}/\textcolor{blue!70!black}{10.62}};
\node[rotate=90, anchor=west, font=\tiny, inner sep=1.2pt]
  at ([xshift=-10.5pt]axis cs:c4,1.74) {\textcolor{black}{1.00}/\textcolor{orange!80!black}{1.22}/\textcolor{blue!70!black}{1.74}};
\node[rotate=90, anchor=west, font=\tiny, inner sep=1.2pt]
  at ([xshift=-3.5pt]axis cs:c4,4.44) {\textcolor{black}{2.69}/\textcolor{orange!80!black}{3.18}/\textcolor{blue!70!black}{4.44}};
\node[rotate=90, anchor=west, font=\tiny, inner sep=1.2pt]
  at ([xshift=3.5pt]axis cs:c4,6.45) {\textcolor{black}{3.91}/\textcolor{orange!80!black}{4.41}/\textcolor{blue!70!black}{6.45}};
\node[rotate=90, anchor=west, font=\tiny, inner sep=1.2pt]
  at ([xshift=10.5pt]axis cs:c4,9.41) {\textcolor{black}{6.24}/\textcolor{orange!80!black}{6.39}/\textcolor{blue!70!black}{9.41}};
\node[rotate=90, anchor=west, font=\tiny, inner sep=1.2pt]
  at ([xshift=-10.5pt]axis cs:c5,1.66) {\textcolor{black}{0.76}/\textcolor{orange!80!black}{1.22}/\textcolor{blue!70!black}{1.66}};
\node[rotate=90, anchor=west, font=\tiny, inner sep=1.2pt]
  at ([xshift=-3.5pt]axis cs:c5,4.17) {\textcolor{black}{2.19}/\textcolor{orange!80!black}{3.18}/\textcolor{blue!70!black}{4.17}};
\node[rotate=90, anchor=west, font=\tiny, inner sep=1.2pt]
  at ([xshift=3.5pt]axis cs:c5,6.14) {\textcolor{black}{3.11}/\textcolor{orange!80!black}{4.41}/\textcolor{blue!70!black}{6.14}};
\node[rotate=90, anchor=west, font=\tiny, inner sep=1.2pt]
  at ([xshift=10.5pt]axis cs:c5,8.9) {\textcolor{black}{5.10}/\textcolor{orange!80!black}{6.39}/\textcolor{blue!70!black}{8.90}};
\node[font=\tiny, anchor=east] at (axis description cs:-0.005,-0.055) {$k=$};
\node[below=1pt, font=\tiny] at ([xshift=-10.5pt]axis cs:attr,0.5) {1};
\node[below=1pt, font=\tiny] at ([xshift=-3.5pt]axis cs:attr,0.5) {3};
\node[below=1pt, font=\tiny] at ([xshift=3.5pt]axis cs:attr,0.5) {5};
\node[below=1pt, font=\tiny] at ([xshift=10.5pt]axis cs:attr,0.5) {10};
\node[below=1pt, font=\tiny] at ([xshift=-10.5pt]axis cs:p0,0.5) {1};
\node[below=1pt, font=\tiny] at ([xshift=-3.5pt]axis cs:p0,0.5) {3};
\node[below=1pt, font=\tiny] at ([xshift=3.5pt]axis cs:p0,0.5) {5};
\node[below=1pt, font=\tiny] at ([xshift=10.5pt]axis cs:p0,0.5) {10};
\node[below=1pt, font=\tiny] at ([xshift=-10.5pt]axis cs:c1,0.5) {1};
\node[below=1pt, font=\tiny] at ([xshift=-3.5pt]axis cs:c1,0.5) {3};
\node[below=1pt, font=\tiny] at ([xshift=3.5pt]axis cs:c1,0.5) {5};
\node[below=1pt, font=\tiny] at ([xshift=10.5pt]axis cs:c1,0.5) {10};
\node[below=1pt, font=\tiny] at ([xshift=-10.5pt]axis cs:c2,0.5) {1};
\node[below=1pt, font=\tiny] at ([xshift=-3.5pt]axis cs:c2,0.5) {3};
\node[below=1pt, font=\tiny] at ([xshift=3.5pt]axis cs:c2,0.5) {5};
\node[below=1pt, font=\tiny] at ([xshift=10.5pt]axis cs:c2,0.5) {10};
\node[below=1pt, font=\tiny] at ([xshift=-10.5pt]axis cs:c3,0.5) {1};
\node[below=1pt, font=\tiny] at ([xshift=-3.5pt]axis cs:c3,0.5) {3};
\node[below=1pt, font=\tiny] at ([xshift=3.5pt]axis cs:c3,0.5) {5};
\node[below=1pt, font=\tiny] at ([xshift=10.5pt]axis cs:c3,0.5) {10};
\node[below=1pt, font=\tiny] at ([xshift=-10.5pt]axis cs:c4,0.5) {1};
\node[below=1pt, font=\tiny] at ([xshift=-3.5pt]axis cs:c4,0.5) {3};
\node[below=1pt, font=\tiny] at ([xshift=3.5pt]axis cs:c4,0.5) {5};
\node[below=1pt, font=\tiny] at ([xshift=10.5pt]axis cs:c4,0.5) {10};
\node[below=1pt, font=\tiny] at ([xshift=-10.5pt]axis cs:c5,0.5) {1};
\node[below=1pt, font=\tiny] at ([xshift=-3.5pt]axis cs:c5,0.5) {3};
\node[below=1pt, font=\tiny] at ([xshift=3.5pt]axis cs:c5,0.5) {5};
\node[below=1pt, font=\tiny] at ([xshift=10.5pt]axis cs:c5,0.5) {10};
\end{axis}
\end{tikzpicture}
\caption{\textbf{Guided identifiers are easier to generate at every position and every list
length.} Prefix recall@$k$---the target's prefix through that position appearing in \emph{any} of
the $k$ returned candidates---for all three encodings under one protocol: retrieval time, after
supervised finetuning, on the identical eval slice ($N=10{,}000$ records) against the \textbf{full production catalog of 654--741M identifiers}, with the
serving-known attribute supplied in the prompt to every arm and sampled decoding ($T{=}1.0$, ten
draws per record). Rates are over all records, so the denominators are identical across arms.
Each position reserves four slots,
$k=1,3,5,10$ left to right. \textbf{Note the log scale.} Guiding leads in all $24$ position-$k$
cells. The margin is largest where guiding acts---$17.70\%$ against $4.22\%$ at position~0,
$k{=}1$---and persists through every free residual level, where the encoding is identical across
arms. \emph{attr} is the prepend arm's extra leading token, a position neither other arm has, since
guiding spends a code of the identifier itself on the attribute rather than adding one; prepending
predicts it at $16.18\%$ but is the weakest arm at every \emph{content} position, and its prefixes
must also carry that token. Numbers are in Table~\ref{tab:perlevel}.}
\label{fig:perlevel}
\end{figure}
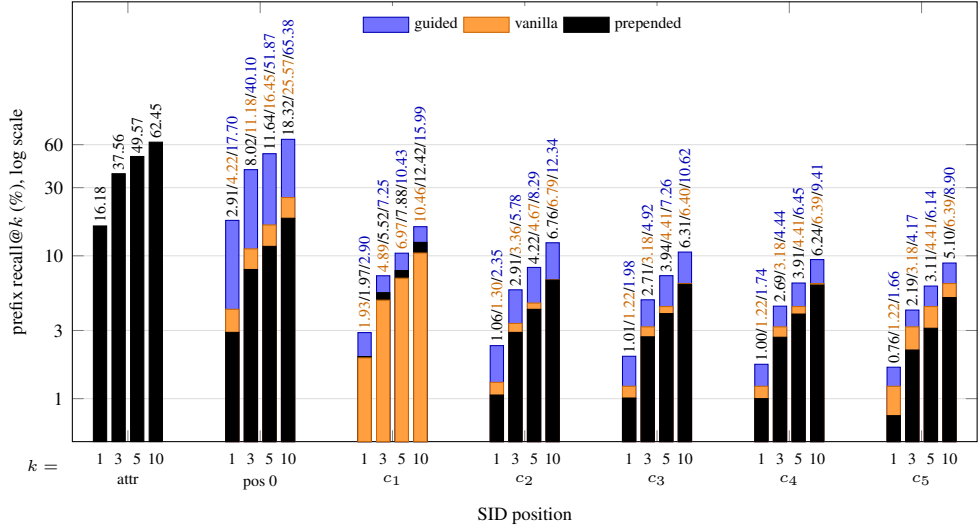

The mechanism claim is that guiding also \emph{scaffolds} the rest of the hierarchy: the residual
levels are unsupervised and identical in construction across arms, so if a semantically coherent
partition sits above them they should become easier to predict in their own right. The per-level
breakdown supports this (Figure~\ref{fig:perlevel}, Table~\ref{tab:perlevel}). Guiding wins the
guided level decisively---$17.70\%$ against $4.22\%$ at $k{=}1$---and it also leads at every
\emph{free residual} level, by $1.50\times$ at $c_1$ and $1.36\times$ at $c_5$, in all $24$
position-$k$ cells. The residual gap is far smaller than the coarse-level one, so the honest
statement is that guiding buys a large improvement where it acts and a modest, consistent one below;
we do not claim the order-of-magnitude residual effect an earlier, smaller-catalog measurement
appeared to show (Appendix~\ref{app:perlevel}). Guiding is nonetheless also a \emph{budget} decision---it buys a far more predictable coarse code
and pays one of six codes for it---and the residual gain has to be weighed against that cost, which
is why we rest the headline claim on end-to-end retrieval rather than on the per-level profile.

Guided assignment grew out of prepending, and the two make structurally
different bets. Prepending adds a symbol \emph{in front of} an unchanged problem: the extra token
narrows the candidate set, but the content identifier behind it is exactly as hard to generate as
before---$c_0$ remains an arbitrary geometric split and the residual levels remain unstructured---so
any gain must come from \emph{conditioning}, the model seeing a useful context token before predicting
an unchanged target. Guiding changes the target instead: the most error-cascading position becomes a
lookup, and every later level is conditioned on a semantically coherent partition rather than a
geometric one. The two accounts predict different residual profiles, which makes the choice testable
rather than rhetorical. Conditioning should improve the code it conditions and otherwise leave
retrieval where the unguided baseline already sits; restructuring should move retrieval itself.

Running a prepended arm through the retrieval protocol of Section~\ref{sec:retrieval}---same base
checkpoint, same recipe, same eval slice, so that all three encodings sit on one axis---separates
them (Table~\ref{tab:retrieval}). Conditioning is real but small: prepending reaches an MRR of
$0.0184$ against the no-attribute baseline's $0.0260$, and at $k{=}1$ it trails it ($0.76$ against
$1.22\%$). Against the full catalog the extra token does not even match leaving the identifier
alone, let alone restructuring it---guiding reaches $1.66\%$ at $k{=}1$ and an MRR of $0.0355$. The attribute therefore helps most when it is
built \emph{into} the identifier rather than placed in front of it.

Two further observations weigh against prepending. It lengthens every identifier, adding a position
at which autoregressive decoding can fail without pruning the content tree, and forcing the
constrained-decoding trie and the vocabulary to be rebuilt for a longer identifier. And it does not
even buy a more legible attribute: prepending predicts its dedicated attribute token at
$16.18\%$ against guiding's $17.70\%$ at the position guiding spends on the same attribute, so the
extra token is no better at carrying the attribute than folding it into the identifier---and it is
the weakest arm at every content position that follows. One caveat on
reading any per-level comparison: the prepended and unguided arms share a content SID and are
compared code-for-code, whereas the guided arm has its own mint and is compared by residual
position; levels are matched by depth in the residual chain, not by identity of codebook.

We are otherwise deliberately candid; the evidence is real but early-stage and industrial.
\textbf{(i) One tested attribute.} Our end-to-end A/B guides on \emph{targeting country}, whose
codebook is head-weighted---the largest bucket holds $22.1\%$ of ads and the ten largest $63.6\%$.
The trie-merge keeps those buckets coherent and the arm performs well regardless, so the occupancy
is workable rather than disqualifying, but a flatter attribute would spread the coarse level further.
Country carries content signal, but indirectly: advertisers tailor creative to the markets they
target, so ads sharing a targeting combination usually address similar markets and a bucket is
correlated with content rather than defined by it. A directly content-bearing attribute would leave
less of the content variance to the residual levels. The most valuable next experiment is
an attribute that is content-bearing, resolvable on its surface, and flatter than country---or a
multi-level guided prefix that spends level~0 on a resolvable constraint and a deeper level on
content once the space has already been narrowed. \textbf{(ii) Single guiding depth evaluated.} The formulation admits an arbitrary guided
prefix $\mathcal{G}$, but every experiment here uses $\mathcal{G}=\{0\}$; the multi-level attribute
hierarchy (e.g., country $\rightarrow$ category $\rightarrow$ subcategory) is untested.
\textbf{(iii) Guided assignment stops at SFT}; no guided RL or deployed serving model. \emph{Future work}: a guiding attribute that is both inference-resolvable and flatter than
country, and multi-level SID that spends level~0 on a resolvable constraint and a deeper level on
content; denser or auxiliary supervision to even out codebook occupancy; composing with
co-occurrence contrastive loss on the free residual levels; and per-attribute constrained decoding
for eligibility.

\section{Conclusion}
We introduced \textbf{Guided SID}: forcing the coarse levels of an RQ-VAE Semantic ID to encode
predefined, interpretable categorical attributes via deterministic supervised index assignment with
learnable codebooks. Guiding the coarse level costs nothing intrinsically (reconstruction even
improves), and in a matched end-to-end A/B it yields large relative gains in autoregressive
retrieval, and a prepended-token control locates that gain in the \emph{restructured coarse code}
rather than in conditioning on the attribute. The gain is driven by a mechanism we make explicit: a supervised coarse partition also makes the
free residual levels easier to generate, though the residual effect is modest beside the
coarse-level one, and it is bought at the cost of one content code.
Guided SID turns the
most load-bearing, most drift-prone part of a Semantic ID---its coarse prefix---into a controllable,
text-grounded, predictable semantic hierarchy.

\subsubsection*{Reproducibility Statement}
The method is specified in full: Section~\ref{sec:method} defines the guided quantisation and the
straight-through estimator, Appendix~\ref{app:codebook} gives the trie-merge that maps an arbitrary
attribute onto a fixed code budget, and Appendix~\ref{app:protocol} states the training recipe, the
evaluation protocol and the candidate-set construction. The experiments run on proprietary industrial
logs that cannot be released; every quantity needed to reproduce the construction on another corpus
is given in the text, and the comparison is internal to the three arms, which share one embedding
snapshot, one attribute labelling, one candidate set and one training recipe. When reproducing on
other corpuses, the guiding feature must be chosen carefully, since the lift ultimately comes from
the information that feature provides.

\subsubsection*{AI use statement}
In this work we used generative AI tools for paper drafting and for experimentation code. We did
not use them for data review, result examination, experimental design, or auditing. We have reviewed
all AI-assisted work: for AI-drafted text in particular, we reviewed each passage and directed
corrections so that the paper states accurately what our research found. We take responsibility for
the final content of this work, including any text, claims or artifacts produced with the aid of
generative AI.

\bibliographystyle{iclr2027_conference}
\bibliography{refs}

\newpage
\appendix

\section{Codebook construction: trie-merge}
\label{app:codebook}
An attribute's values are compressed to exactly $K=256$ codes by a bottom-up trie merge. Each value is
a path in a trie whose sibling nodes share a common prefix (for targeting country, the sorted set of
targeted country codes; for a content taxonomy, the category path), so siblings are the most similar
values.

\begin{verbatim}
build a weighted-order trie over all attribute values   # more popular values first
while num_leaves(trie) > K:
    p <- the parent whose smallest child holds the fewest items
    merge p's two smallest children into one leaf
\end{verbatim}

Large values keep their own code (they are never the smallest sibling); small values merge with their
closest relatives (shared prefix). For targeting country, 247 base codes and $481{,}343$ co-targeting
combinations compress to 256 entries: 124 single-country-prefix entries (21 pure, 103 absorbing small
combinations) and 132 multi-country entries. Buckets are unbalanced only where the data is.
Measured over the $4.42$B-ad universe the encoding populates 241 of the 256 entries: the largest
holds $22.1\%$ of ads ($56.7\times$ the per-code target, a single dominant market), the
second---the ``other/no country'' catch-all---$16.2\%$ ($41.6\times$), the ten largest $63.6\%$
between them, and the median entry $0.15\times$. This head-weighting is why a coarse-level accuracy
is a property of the attribute's real distribution, and is the main limitation of the construction:
a code that absorbs a fifth of the corpus carries correspondingly little information
(Section~\ref{sec:limitations}).

\section{Full intrinsic tables}
\label{app:intrinsic}
Cluster precision / separation recall of the guided (country-at-L0) tokenizer against several
ground-truth attributes, at the coarse prefix (L1, country only) and the full SID
(Table~\ref{tab:appB1}). Category ground truth is a content taxonomy; country ground truth is a
\emph{proxy} (content-inferred), so its L1 recall (0.892) understates the forced encoding, which is
1.0 against the true targeting label by construction.

\begin{table}[t]
\centering
\caption{Cluster precision / separation recall by ground-truth attribute, at the coarse prefix (L1)
and the full SID (L1--L6). Guided tokenizer.}
\label{tab:appB1}
\begin{tabular}{lcccc}
\toprule
 & \multicolumn{2}{c}{L1 (coarse = country)} & \multicolumn{2}{c}{Full SID (L1--L6)} \\
\cmidrule(lr){2-3}\cmidrule(lr){4-5}
Ground-truth attribute & Prec. & Recall & Prec. & Recall \\
\midrule
Category (L1) & 0.414 & 0.641 & 0.956 & 0.626 \\
Category (L2) & 0.079 & 0.953 & 0.920 & 0.937 \\
Category (L3) & 0.053 & 0.969 & 0.904 & 0.959 \\
Category (L4) & 0.044 & 0.973 & 0.894 & 0.969 \\
Country (proxy) & 0.384 & 0.892 & 0.845 & 0.793 \\
Language & 0.634 & 0.810 & 0.988 & 0.717 \\
Objective & 0.370 & 0.757 & 0.983 & 0.547 \\
\bottomrule
\end{tabular}
\end{table}

\begin{table}[t]
\centering
\caption{Country precision / recall by SID depth (guided tokenizer). Recall stays high from L1
(forced); precision grows as deeper levels add specificity.}
\label{tab:appB2}
\begin{tabular}{lcc}
\toprule
Prefix & Precision & Recall \\
\midrule
L1 & 0.384 & 0.892 \\
L1L2 & 0.459 & 0.892 \\
L1L2L3 & 0.555 & 0.924 \\
L1--L4 & 0.745 & 0.876 \\
L1--L5 & 0.813 & 0.826 \\
L1--L6 & 0.845 & 0.793 \\
\bottomrule
\end{tabular}
\end{table}

\section{End-to-end A/B protocol}
\label{app:protocol}
Both SIDs are minted together from the same content embeddings and joined 1:1, so coverage is
${\sim}$identical (${\sim}100\%$ over the minted item universe) and the only variable is the encoding.
Over the interacted (history) item set, coverage is ${\sim}98.5$--$99\%$ for \emph{all} arms---a
shared embedding-coverage ceiling, not a guided-specific deficit. A record is kept only if every
one of its items resolves in every arm's codebook, which keeps histories contiguous and makes the
rendered datasets identical across arms; the retained records average $19.3$ items. All three arms share the CPT recipe (step 20K) then SFT
(step 10K,
user-interaction-history $\rightarrow$ next item). Evaluation: $N=10{,}000$ per arm,
sampled decoding at $T{=}1.0$ with ten draws per record, constrained
decoding over a trie spanning the \emph{full} production catalog---654M identifiers for the
baseline, 741M for the prepended arm and 655M for the guided arm, every ad in the corpus rather than
a sampled subset---so no candidate-set construction enters the comparison, the candidate set is
held fixed across arms, and only the encoding varies. Every evaluation target is reachable in every
arm's trie by construction. Smaller candidate sets, built from the evaluation targets plus further
items drawn from the training pool, were also measured and are reported in
Appendix~\ref{app:perlevel}.
recall@$k$ is an exact full-identifier match anywhere in the returned list; country\_match@1 asks
whether the retrieved item's coarse attribute matches the target's. The attribute the arms are
guided on is known to the serving system before a candidate is chosen, so it is supplied in the
prompt to every arm, including the baseline.

\section{The prepended-token encoding}
\label{app:prepend}
The third arm exposes the same attribute to the retriever without touching the content codebook.
Rather than assigning the attribute to a quantization level, it leaves the tokenizer's content
codebooks alone and emits the attribute as an extra leading token,
$\langle a(x)\rangle\,c_0 \dots c_{L-1}$, so the identifier is $L{+}1$ tokens instead of $L$ and the
model narrows to the attribute before generating an \emph{unchanged} content SID. Concretely, with
the six content codes of our setup an item is written
\texttt{<TOKEN><country/$g$><ad\_0/$c_0$>$\dots$<ad\_5/$c_5$></TOKEN>}, where $g$ is the attribute
code and $c_0\dots c_5$ are exactly the codes the unguided baseline assigns to that item---the
prepended and unguided arms share a content SID by construction, which is what lets us compare them
code-for-code.

This is the non-invasive option, and for purely \emph{extrinsic} attributes---delivery constraints
such as age or gender, which describe who may see an item and say little about what it is---it is
arguably the more principled place to put the signal. Targeting country sits between the two: it
constrains delivery, but because advertisers specialize by market it also correlates with content,
which is what makes it viable to guide on. It is also where guided assignment came from:
we arrived at guiding by asking what would happen if, instead of adding a symbol \emph{in front of}
the identifier, the attribute became part of the identifier itself. The costs are that it lengthens
every identifier, adds one more position at which autoregressive decoding can fail without pruning
the content tree, extends the vocabulary by one code block, and---because constrained decoding walks
a prefix trie over SIDs---requires that trie and its per-level offsets to be rebuilt for the longer
identifier. The two are complementary rather than exclusive: one can prepend extrinsic eligibility
tokens \emph{and} guide the coarse content levels.

\section{Per-level accuracy and base-rate baselines}
\label{app:perlevel}

The main text rests on end-to-end retrieval (Table~\ref{tab:retrieval}). This appendix gives the
per-level breakdown behind it (Table~\ref{tab:perlevel}) and one caveat on how to read it.

\begin{table}[t]
\centering
\caption{Prefix recall@$k$ (\%) by SID position, full catalog (654--741M identifiers),
$N=10{,}000$ records, sampled decoding at $T{=}1.0$ with ten draws.
The entry at position $\ell$ and list length $k$ is the fraction of records whose target prefix
through $\ell$ appears in \emph{any} of the $k$ returned candidates; the $c_5$ row is therefore the
full-identifier recall@$k$ of Table~\ref{tab:retrieval}. The prepended arm is matched on
\emph{content}: its content position $\ell$ is read off position $\ell+1$ of its seven-position
identifier, so its prefix at any depth must also carry the attribute token right. Guiding leads
every one of the $24$ position-$k$ cells.}
\label{tab:perlevel}
\small
\begin{tabular}{lcccccc}
\toprule
Arm & pos~0 & $c_1$ & $c_2$ & $c_3$ & $c_4$ & $c_5$ \\
\midrule
\multicolumn{7}{l}{\emph{$k=1$}\quad\scriptsize(prepend attribute token: 16.18)} \\
\quad Guided & \textbf{17.70} & \textbf{2.90} & \textbf{2.35} & \textbf{1.98} & \textbf{1.74} & \textbf{1.66} \\
\quad Vanilla & 4.22 & 1.93 & 1.30 & 1.22 & 1.22 & 1.22 \\
\quad Prepended & 2.91 & 1.97 & 1.06 & 1.01 & 1.00 & 0.76 \\
\addlinespace
\multicolumn{7}{l}{\emph{$k=3$}\quad\scriptsize(prepend attribute token: 37.56)} \\
\quad Guided & \textbf{40.10} & \textbf{7.25} & \textbf{5.78} & \textbf{4.92} & \textbf{4.44} & \textbf{4.17} \\
\quad Vanilla & 11.18 & 4.89 & 3.36 & 3.18 & 3.18 & 3.18 \\
\quad Prepended & 8.02 & 5.52 & 2.91 & 2.71 & 2.69 & 2.19 \\
\addlinespace
\multicolumn{7}{l}{\emph{$k=5$}\quad\scriptsize(prepend attribute token: 49.57)} \\
\quad Guided & \textbf{51.87} & \textbf{10.43} & \textbf{8.29} & \textbf{7.26} & \textbf{6.45} & \textbf{6.14} \\
\quad Vanilla & 16.45 & 6.97 & 4.67 & 4.41 & 4.41 & 4.41 \\
\quad Prepended & 11.64 & 7.88 & 4.22 & 3.94 & 3.91 & 3.11 \\
\addlinespace
\multicolumn{7}{l}{\emph{$k=10$}\quad\scriptsize(prepend attribute token: 62.45)} \\
\quad Guided & \textbf{65.38} & \textbf{15.99} & \textbf{12.34} & \textbf{10.62} & \textbf{9.41} & \textbf{8.90} \\
\quad Vanilla & 25.57 & 10.46 & 6.79 & 6.40 & 6.39 & 6.39 \\
\quad Prepended & 18.32 & 12.42 & 6.76 & 6.31 & 6.24 & 5.10 \\
\bottomrule
\end{tabular}
\end{table}

\end{document}